\documentclass{iaus}
\usepackage{graphicx}

\title[The case for a close-in perturber to GJ 436 b]
{The case for a close-in perturber to GJ 436 b}

\author[Ribas et al.]   
{Ignasi Ribas$^{1,3}$, Andreu Font-Ribera$^{1,3}$, Jean-Philippe
Beaulieu$^{2,3}$, Juan Carlos Morales$^1$, \and Enrique Garc\'{\i}a-Melendo$^4$}
\affiliation{$^1$Institut de Ci\`encies de l'Espai (CSIC-IEEC), Campus UAB,
08193 Bellaterra, Spain
\\[\affilskip]
$^2$Institut d'Astrophysique de Paris, CNRS (UMR 7095), Paris, France
\\[\affilskip]
$^3$The HOLMES collaboration
\\[\affilskip]
$^4$Esteve Duran Observatory Foundation, Montseny 46, 08553 Seva, Spain}

\pubyear{2008}
\volume{253}  
\pagerange{}
\jname{Transiting Planets}
\editors{TBD}
\begin{document}

\maketitle

\begin{abstract}
The increasing number of transiting planets raises the possibility of 
finding changes in their transit time, duration and depth that could be 
indicative of further planets in the system. Experience from eclipsing 
binaries indeed shows that such changes may be expected. A first obvious 
candidate to look for a perturbing planet is GJ 436, which hosts a hot 
transiting Neptune-mass planet in an eccentric orbit. Ribas et al. (2008) 
suggested that such eccentricity and a possible change in the orbital 
inclination might be due to a perturbing small planet in a close-in orbit. 
A radial velocity signal of a 5~M$_\oplus$ planet close to the 2:1 
mean-motion resonance seemed to provide the perfect candidate. Recent new 
radial velocities have deemed such signal spurious. Here we put all the 
available information in context and we evaluate the possibility of a 
small perturber to GJ~436~b to explain its eccentricity and possible 
inclination change. In particular, we discuss the constraints provided by 
the transit time variation data. We conclude that, given the current data, 
the close-in perturber scenario still offers a plausible explanation to 
the observed orbital and physical properties of GJ~436~b.
\end{abstract}

\section{Introduction}

The study of eclipsing binary systems was initiated by the discovery by 
John Goodricke in 1783 and subsequent proof by Edward C. Pickering in 1881 
that the variable star Algol is indeed composed of two stars undergoing 
eclipses. In the last century, the study of eclipsing binary systems has 
been a major component to Stellar Astrophysics as a whole, by improving 
our understanding of a variety of phenomena in binaries, by providing 
valuable information on the structure and evolution of stars, and by 
serving as indicators of, e.g., age or distance (see, e.g., Andersen 1991; 
Guinan 1993; Hilditch 2001; Ribas 2006; Bonanos 2007). Eclipsing binaries 
still continue to play an important role in this respect. However, in 
recent years we have witnessed a rebirth of the study of eclipsing binary 
systems in the particular case of very unequal mass and brightness ratios, 
i.e., transiting planets. Being eclipsing binaries a field of long 
tradition (both observational and theoretical) it is convenient to use 
such outstanding background and adapt it to suit current needs.

A particular aspect of eclipsing binary research has been the study of 
detached systems that show variable light curves. In some cases, the 
origin of the variability can be traced to the appearance, migration, and 
disappearance of inhomogeneities (starspots) on the surface of one or both 
components but in some instances the observed changes are of a more 
fundamental nature. For example, the eclipsing binaries SS~Lac and SV~Gem 
stopped eclipsing some 60 years ago (Torres \& Stefanik 2000; Guilbault et 
al. 2001), V906~Sco stopped eclipsing in 1918, then restarted in 1963 and 
stopped again in 1986 (Lacy et al. 1999), and IU~Aur shows a fast 
variation of eclipse depth (Drechsel et al. 1994). All these changes are 
thought to originate from the perturbations of a third star in the system 
(S\"oderhjelm 1975; Mazeh \& Shaham 1976). Obviously, such large 
perturbations are not very common among eclipsing binaries but it is also 
true that not many systems have been observed intensively enough and with 
a sufficiently long time baseline to uncover slow light curve variations.

In the case of transiting planets, variations in light curve properties 
have been predicted to occur from various sources, most notably from the 
effect of perturbing further bodies in the system. Transit time variations 
have been the subject of intense attention and indeed been proposed as a 
way to detect smaller perturbing planets (Miralda-Escud\'e 2002; Holman \& 
Murray 2005; Agol et al. 2005). But not only the transit central time may 
suffer variations. The duration and depth of the transit (Schneider 1994; 
Miralda-Escud\'e 2002; Laughlin et al. 2005) may also be modified because 
of changes in the orbital inclination, semi-major axis, eccentricity and 
argument of periastron.

The numerous ongoing surveys from both the ground and space are now 
producing new transiting planet discoveries at an ever increasing pace. 
The total tally of exoplanets undergoing transit events has already 
surpassed 50. With these increasing statistics and the eclipsing binary 
experience one may wonder if changes in the light curve are already 
observable in some cases.

\section{GJ 436}

The M2.5-dwarf GJ~436 was discovered to host a Neptune-mass planet 
(22~M$_\oplus$) in a 2.6-d orbit by Butler et al. (2004). Two properties 
made this object especially interesting, namely its relatively small mass 
and a surprising non-zero eccentricity of about 0.15. Such value of the 
eccentricity was recently confirmed by the analysis of Maness et al. 
(2007), hereafter M07. Butler et al. (2004) also obtained high-precision 
photometry and ruled out the possibility of a transit with a depth greater 
than 0.4\%. However, a surprise came with the actual detection of transits 
from GJ~436~b with a depth of 0.7\% by Gillon et al. (2007b), thus 
becoming, by far, the smallest transiting planet yet detected. A series of 
studies, mostly using Spitzer, have greatly contributed to establishing 
the properties of the planet and also to strengthen the case for an 
eccentric orbit by observing the occultation event at orbital phase 0.59 
(Deming et al. 2007; Gillon et al. 2007a).

The origin of the high eccentricity of GJ~436~b was investigated in detail 
by M07 and Demory et al. (2007). Both studies conclude that the 
circularization timescale ($\sim$10$^8$~yr) is significantly smaller than 
the old age of the system ($\gtrsim$6$\cdot$10$^9$~yr) when assuming 
reasonable values for the planet's tidal dissipation parameter. M07 also 
pointed out the presence of a long-term trend with a value of 
1.3~m~s$^{-1}$ per year on the systemic radial velocity of GJ~436. Thus, 
the authors investigated the possibility that the eccentricity and the 
long-term velocity trend could be explained from the perturbation exerted 
by an object in a wider orbit without reaching conclusive results.

But GJ~436~b has yet another remarkable trait and this is the near-grazing 
nature of its transit. The impact parameter of the transit was found to be 
about 0.85, which implies an orbital inclination of 86.3$^\circ$. If the 
inclination happened to be just 85.3$^\circ$ the planet would not cross 
the disk of the star. GJ~436~b makes an ideal system to find evidence for 
a perturbing small planet, because of the telltale non-zero eccentricity, 
but also to put severe constraints on the properties of the perturber 
owing to the extreme sensitivity of the current configuration to small 
changes in the orbital inclination. In Ribas et al. (2008), hereafter 
RFB08, we proposed an alternative possibility to explain the eccentricity 
of GJ~436~b, namely the perturbation from a relatively small planet in a 
close-in orbit. Our scenario is described in the next section.

\section{A second planet around GJ 436?}

\begin{figure}
\begin{center}
 \includegraphics[width=9cm]{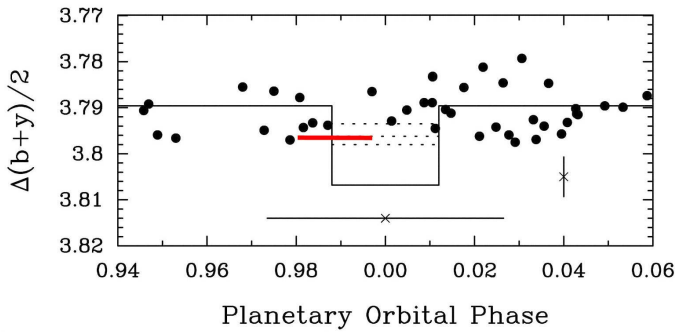}
 \caption{Photometry of GJ~436 from Butler et al. (2004) with the true 
depth of the transit marked by the thick line between phase $\sim$0.98 and 
phase $\sim$0.00. Several measurements fall inside the transit window. 
Figure adapted from Butler et al. (2004).}
   \label{fig:trans}
\end{center}
\end{figure}

A possible explanation for the apparently contradicting results concerning 
the detection of transits is that the orbital inclination has indeed 
changed during the 3.3-year interval between the different photometric 
observations. Calculations show that an orbital inclination 
$\lesssim$86$^{\circ}$ would have made the transit undetectable to the 
photometric measurements of Butler et al. (2004). From these 
considerations a small variation of the inclination angle at a rate of 
roughly $\sim$0.1$^{\circ}$~yr$^{-1}$ could make both the Butler et al. 
(2004) non detection and Gillon et al. (2007b)'s discovery of transits 
compatible. Note that this is only a possible scenario since the 
photometry of Butler et al. has relatively sparse phase coverage. As can 
be seen in Fig. \ref{fig:trans}, several measurements should have betrayed 
the presence of the transit, although with low significance.

For more accurate estimates we carried out direct integrations of the 
equations of motion using the Mercury package (Chambers 1999). We started 
with an inner planet in a circular orbit and with the currently observed 
semi-major axis. Then, we considered different combinations of mass (from 
1 to 14~M$_{\oplus}$), semi-major axis (from 0.04 to 0.1 AU), eccentricity 
(from 0.05 to 0.3) and inclination (from 85$^{\circ}$ to 45$^{\circ}$) for 
the perturber. The integrations were performed for a time interval of 
10$^5$ yr to guarantee the stability of the planetary systems.  We further 
explored semi-major axis values at mean-motion resonances (MMRs). Location 
in a MMR can be a stabilizing factor and also perturbations can reach 
their maximum efficiency (e.g., Agol et al. 2005). Integrations for 
semi-major axes corresponding to the following MMRs were carried out: 3:2, 
5:3, 2:1, 3:1, and 4:1. In all cases, the presence of the planet in a MMR 
increased the stability and, further, perturbing planets with smaller 
masses were able to induce the observed eccentricity and orbital 
inclination change to the inner planet. For the strongest 2:1 resonance we 
found a lower limit to the perturbing planet mass of only 1~M$_{\oplus}$ 
at an extreme eccentricity and relative inclination. For the general case 
of a perturbing planet with 3--7~M$_{\oplus}$, eccentricity values of 
0.15--0.20 and initial inclination differences of only 5--15$^\circ$ were 
sufficient to explain the observed eccentricity and rate of inclination 
change of the inner planet.

In the analysis we neglected tidal dissipation since we focused on the 
current snapshot of the orbital configuration of the system but the 
planets must be undergoing significant tidal dissipation because of the 
non-zero eccentricity. Other effects have been neglected at this stage, 
which include precession caused by the quadrupole moment of the star and 
by General Relativity (GR).

Further, in RFB08 we carried out a re-analysis of the available radial 
velocity data on GJ 436 and identified a second peak (of quite low 
significance) on the periodogram with a period of 5.18 d. Such peak 
corresponded to a planet with a minimum mass of 4.7~M$_{\oplus}$ and close 
to the 2:1 MMR with the inner planet. Remarkably, a planet of such 
characteristics would be a perfect match to the perturbing object revealed 
by the evidence on the orbital eccentricity and inclination change.

\section{Discussion}

During the Symposium, further radial velocity data on GJ 436 were 
presented by two groups (Howard, Bonfils, this volume) and the 5.18 d peak 
is not present. The amount and the accuracy of the new data is also 
superior and the authors do not find any further significant signals above 
the noise level. Thus, it is now clear that the planet proposed by RFB08 
to be responsible for the observed perturbations comes from a spurious 
signal. In addition, the new velocities show the long-term velocity trend 
to be an artifact from insufficient time baseline and sparse coverage. But 
independently of the precise identity of the perturbing planet, one can 
constrain its properties by measuring the rate of inclination change and 
also the presence of transit time variations. Obviously, GJ 436 has been 
the focus of attention and numerous observations have been acquired during 
this season, as illustrated from different papers presented in the 
Symposium.

Conclusive evidence of the existence of a perturber would come from the 
measurement of variations in the transit shape. This cannot be done 
directly by comparing inclination values from different studies because of 
correlations with other parameters. The best way is to look for changes in 
the total transit duration, which is a fundamental quantity derived from 
the photometry. Besides the 2007 season data, new photometry has been 
presented by Alonso et al. (2008). From a ground-based transit of 
outstanding quality the authors measure a marginally longer transit 
duration than in 2007, which can be translated into a rate of inclination 
change of +0.03$\pm$0.05$^\circ$~yr$^{-1}$. This is both compatible with 
zero and with the $\sim$0.1$^\circ$~yr$^{-1}$ rate suggested by RFB08. New 
transit data have been acquired by HST (Bean et al. 2008) in Dec'07 and 
Jan'08 but the transit duration information is not given, and also the 
data do not enhance the time coverage. Transit duration measurements from 
amateur astronomers\footnote{\tt 
http://brucegary.net/AXA/GJ436/gj436.htm}, while extending the current 
time baseline, do not have sufficient accuracy for a current estimate of 
the possible change (some 1--2 min compared with scatter of 5--10 min). 
High-precision observations during the coming seasons will permit the 
measurement of the putative transit duration variation and, if confirmed, 
to put stringent constrains on the perturbing planet.

Another way to test the presence of perturbing planets is via transit time 
variations (TTVs). When two planets are near an MMR, their interaction 
gives rise to libration motions that translate into relatively large 
variations in the time of conjunction (i.e., transit) with typical short 
characteristic timescales (months). The detection of TTVs is a clear and 
unambiguous signal of further planets in a system. However, there are two 
very important caveats. One is that the inverse problem may not be well 
defined. In other words, it is not guaranteed that there exists a unique 
combination of planet parameters that will reproduce a given TTV signal. 
Very high timing accuracy may be needed to disentangle subtle differences 
but there may be some inherent degeneracies, especially if time coverage 
of only a few years is available.

\begin{figure}
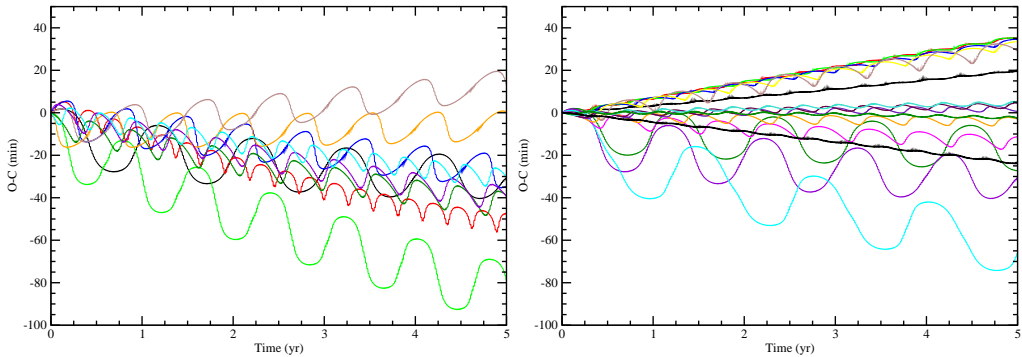
 \begin{center}
 \includegraphics[width=6.7cm]{ribas_fig2a.eps}
 \includegraphics[width=6.7cm]{ribas_fig2b.eps}
 \caption{TTVs arising from a perturber to GJ~436~b. The left panel shows 
TTVs from a number of configurations inside the 1-$\sigma$ uncertainties 
of the perturber planet in RFB08. The right panel depicts TTVs from the 
nominal parameters of the perturber but from different values of the 
longitude of the line of nodes, which is an orbital element unconstrained 
from the radial velocity data.}
   \label{fig:ttvs}
\end{center}
\end{figure}

The second caveat concerns the opposite situation, and thus it is more 
relevant here. We would like to stress that the lack of a TTV signal, in 
general, does not rule out the presence of a planet inferred from, e.g., 
radial velocities. This is because some of the orbital elements are not 
constrained by the radial velocity data and these may be very relevant to 
TTVs. Common practice is that of averaging over unknown elements but, 
while statistically sound, this does not apply to a single studied case. 
In addition, TTV signals may vary quite significantly with small changes 
of orbital elements. All this is illustrated in Fig. \ref{fig:ttvs} and 
also by Veras (this volume). For example, very small TTVs can be found for 
certain configurations at the center of the strong 2:1 MMR. Therefore, the 
lack of a TTV cannot be generally used as a strong proof against the 
presence of a perturbing planet. In the case of GJ 436, a statement like 
``there is no further planet because we do not see the expected TTVs of 
the order of minutes'' is not strictly correct given the limited orbital 
constraints we have.

\begin{figure}
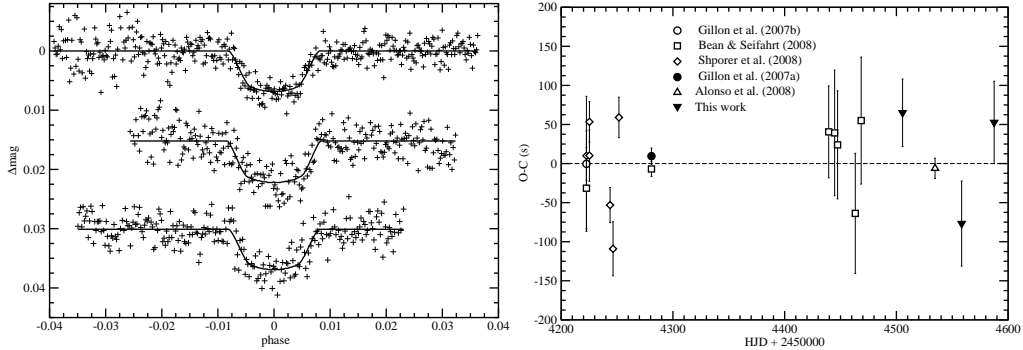

\begin{center}
 \includegraphics[width=6.7cm]{ribas_fig3a.eps}
 \includegraphics[width=6.7cm]{ribas_fig3b.eps}
 \caption{{\em Left}: Three transit events observed with the 60-cm 
telescope of Esteve Duran Observatory and the corresponding best fits. 
{\em Right}: O-C diagram for the published timing data plus our three new 
measurements.}
   \label{fig:timing}
\end{center}
\end{figure}

Transit timing measurements of GJ~436~b have been published so far by 
Alonso et al. (2008), Shporer et al. (2008), and Bean et al. (2008), and 
measurements have also been presented by Winn and Demory (both in this 
volume), and by amateurs.  We have also carried out our own measurements 
and we have observed three transits from the 60-cm telescope at Esteve 
Duran Observatory. As can be seen in Fig. \ref{fig:timing}, the rms of the 
photometry is of the order of 1.5--2 mmag. The three transit mid-time 
measurements are: HDJ2454505.51379$\pm$0.00050, 
HJD2454558.39010$\pm$0.00063, and HJD2454587.47447$\pm$0.00061; with total 
transit durations (in min) of 60.9$\pm$1.1, 63.0$\pm$1.5, and 
60.6$\pm$1.4, respectively. A weighted least squares fit yields the 
following ephemeris: $T_{\rm mid} = \mbox{HJD}2454280.78167(11) + 
2.6438975(15)$; with a $\chi^2$ value of 1.8. The O-C residuals from the 
linear ephemeris are given in Fig. \ref{fig:timing} for all the published 
timings. From the data available it is still early to draw conclusions, 
but the timings do not reveal significant variations from a linear trend. 
At this point, any possible modulation should be below $\sim$1 minute. 
This does not favour (although it strictly does not rule out) a perturbing 
planet in MMR.

Very recently, Bean \& Seifahrt (2008) have re-analyzed the radial 
velocities of M07 and used the available timing measurements by 
considering a sophisticated model with planet-planet interactions. The 
authors come up with a possible perturber with a mass of 5~M$_{\oplus}$ at 
0.043~AU. This is close to the planet proposed by RFB08 but located 
outside the 2:1 MMR because of the additional TTV constraints. Such 
solution has a significance not much greater than other mass vs. 
semi-major axis combinations. In light of their simulations, the authors 
conclude that a close in perturber is possible.

The case for a 5~M$_{\oplus}$ at 2:1 MMR has also been studied by Mardling 
(2008). In this case, the model considered includes the tidal interactions 
between the planets. The analysis indicates that the precise configuration 
proposed is stable but that it would not stop the inner planet from being 
circularized at some point. The allowed region in the mass vs. semi-major 
axis plane for a perturbing planet will have to be defined with all the 
observational constraints, plus the orbital perturbation model and tidal 
energy dissipation. We plan to do so in the near future.

\section{Conclusions}

New radial velocity data have ruled out the presence of a 5~M$_\oplus$ 
planet with a period of 5.18 d in GJ~436. Although the new data set a more 
stringent limit to possible further planet in GJ~436, the scenario of a 
close-in perturber to GJ~436~b is still plausible. Strong proof should 
come from changes in the transit duration measured over the coming 
seasons. If such changes are there, this will nail the case for a 
perturbing object in a slightly non-coplanar orbit, much in the same way 
as for some eclipsing stellar binaries. This will give rise to the 
interesting concept of ``transient transits''. Still, it is possible that 
the duration is stable over time and this would rule out a perturber in 
mild non-coplanarity.

The most certain observational fact is the eccentricity of GJ~436~b's 
orbit. Besides the close-in perturber scenario, there are a number of 
other possible explanations. A distant perturber was proposed by M07 and 
also by Bonfils (this volume). This would explain the radial velocity 
trend and be responsible for the eccentricity pumping. However, the trend 
seems not to be confirmed by newer data and, further, the effect of GR 
precession may prevent the building of significant eccentricity. This is 
because the GR timescale for GJ~436~b is only 15\,000 yr and any 
eccentricity pumping effect with a longer timescale will not be efficient. 
The other obvious scenario is that of a large value of the tidal 
dissipation constant $Q'_{\rm p}$ for GJ~436~b. The value needed is 
10$^{6-7}$, which is one to two orders of magnitudes larger than Neptune's 
and larger than that of any object in the Solar System. A large value of 
$Q'_{\rm p}$, if it can be generalized, should also be made compatible 
with the distribution of eccentricities of close-in planets. Further, 
there is weak evidence of tidal heating on GJ~436~b (Deming et al. 2007), 
which advocates for a normal $Q'_{\rm p}$. Finally, one may also think of 
a recent event (less than 100 Myr ago) that pumped up the eccentricity of 
the planet (as suggested by Zakamska \& Tremaine 2004, in another 
context), such as the close passing of a star or massive object. However, 
at this point this is just mere speculation.

We believe that a close-in perturber is still the most likely scenario to 
explain the observations of GJ~436. Some of the small planets found in 
around mass stars seem to belong to multiple systems, such as Gl~876, 
Gl~581, or the recent discovery of HD 40307. Thus, given this remarkable 
parallelism, it would not be a surprise if GJ~436 hosts more than one 
planet. This would also be the case in the framework of the hypothesis of 
packed planetary systems of Barnes \& Raymond (2004). As more transiting 
planets are discovered, the chances of observing variations in their 
transit properties with time increase. The case of near-grazing events is 
especially suitable because of their sensitivity to perturbers. In the 
coming years, this technique combined by intensive studies of transiting 
planets (ensured by the interest in the field) should provide us with new 
insight into the architecture of planetary systems.

\end{document}